\documentclass[aps,prl,twocolumn,groupedaddress]{revtex4}
\usepackage{graphicx}

\begin{document}

\title{Self-optimization of optical confinement in ultraviolet photonic crystal slab laser}
\author{A. Yamilov$^{1,2}$, X. Wu$^{1}$, X. Liu$^{3}$, R. P. H. Chang$^{3}$ and H. Cao$^{1}$}
\affiliation{$^{1}$ Department of Physics and Astronomy, Northwestern University, Evanston, Illinois, 60208\\
$^{2}$ Department of Physics, University of Missouri-Rolla, Rolla, MO 65409\\
$^{3}$ Department of Materials Science and Engineering, Northwestern University, Evanston, Illinois, 60208}
\email{yamilov@umr.edu; h-cao@northwestern.edu}

\begin{abstract}
We studied numerically and experimentally the effects of structural disorder on the performance of ultraviolet photonic crystal slab lasers. Optical gain selectively amplifies the high-quality modes of the passive system. For these modes, the in-plane and out-of-plane leakage rates may be automatically balanced in the presence of disorder. The spontaneous optimization of in-plane and out-of-plane confinement of light in a photonic crystal slab may lead to a reduction of the lasing threshold. 
\end{abstract}
\maketitle

Over the past few years, tremendous progress has been made in design and fabrication of photonic crystal slab (PhCS) lasers that operate in infra-red (IR) spectrum range \cite{mode_delocalization,Painter_noda_park,noda_science,Akahane:2003,akahane_nat_mat,scherer2,bandedge_exp}. To realize an ultraviolet (UV) PhCS laser, the feature size has to be reduced roughly by a factor of four \cite{our_apl,our_pbg_calc} compared to IR PhCS. Fabrication of such fine structures inevitably generates random deviations from the perfectly ordered structures. In passive PhC devices such uncontrollable disorder has detrimental effect as it contributes to optical losses and limits light propagation length. However, it is not clear how the disorder would affect the performance of an {\it active} device, e.g., a PhCS laser. In this letter we illustrate, through numerical simulation and experiment, that structural disorder in a PhC laser may not be as detrimental as in a PhC waveguide. Under some circumstances structural disorder enables self-optimization of optical confinement in a PhCS, leading to a reduction of lasing threshold.

A PhCS utilizes index guiding to confine light to the plane of the slab \cite{phcs_confinement}. In-plane confinement is realized either via a defect state located inside a photonic bandgap (PBG) \cite{Painter_noda_park,Akahane:2003,akahane_nat_mat,scherer2}, or a bandedge state with vanishing group velocity \cite{polymer_phcs,gain_at_be,noda_science,bandedge_exp}. Light may escape from the PhCS vertically through the top/bottom interfaces into air/substrate or laterally via the edge of the periodic pattern into air or unpatterned part of the slab. The vertical leakage rate is characterized by the out-of-plane energy loss per optical cycle $Q_{\perp}^{-1}$, and the lateral by $Q_{||}^{-1}$. A defect state spatially localized in the vicinity of an intentionally-introduced structural defect typically has large leakage in the vertical direction, i.e., $Q_{\perp}^{-1} \gg  Q_{||}^{-1}$. For a bandedge state, the lateral leakage usually dominates over the vertical one, $Q_{||}^{-1} \gg  Q_{\perp}^{-1}$. The total loss is described by $Q_{tot}^{-1}=Q_{\perp}^{-1}+Q_{||}^{-1}$. Low lasing threshold demands maximization of $Q_{tot}$, which is hindered by $Q_{\perp}$ for a defect state and $Q_{||}$ for a bandedge state. Several designs aim at optimization of PhCS lasers by balancing $Q_{\perp}$ and $Q_{||}$ via ``gentle localization'' \cite{Akahane:2003}, e.g., phase-slip\cite{phase_slip,scherer2}, double-heterostructure \cite{akahane_nat_mat}.

Recently we realized the first UV PhCS laser \cite{our_apl}. ZnO films were grown on sapphire substrates by plasma enhanced MOCVD \cite{xiang_liu}. Hexagonal arrays of cylindrical air voids were patterned in the ZnO films by focused ion beam (FIB) etching technique. The lattice constant $a \sim 120$ nm, the radius of air cylinders $R \sim 30$ nm. Post thermal annealing was employed to remove the FIB damage. Single-mode lasing at room temperature was realized with optical pumping. The scanning electron micrograph (SEM) of a ZnO PhCS is shown in Fig. 1(a). Despite the long-range periodicity exhibited in the inset of Fig. 1(a), Fig. 1(b) reveals the deviation of the fabricated pattern from the ideal honeycomb structure. Such  ``crescent'' deviation \cite{vos_disorder} caused optical scattering on the length scale of a few lattice constants. It was  expected to enhance radiative leakage of a PhCS laser based on either defect state or bandedge mode. Moreover, the propagation loss in a passive PhCS caused by random\cite{disorder_latest} scattering was predicted to increase dramatically near a photonic bandedge \cite{disorder_at_be}, where the bandedge-type PhCS laser operates. Despite of these pessimistic expectations based on passive systems, we show that the performance of a PhCS laser may be less susceptible to the detrimental effects of structural disorder. This is because optical gain predominantly amplifies the mode with the highest quality factor $Q_{tot}$. For the highest-$Q_{tot}$ mode, the vertical and lateral leakage rates may be automatically balanced in the presence of disorder. This implies that an appropriate amount of structural disorder could lead to spontaneous optimization of in-plane and out-of-plane confinement of light in a PhCS. 

\begin{figure}
\vskip 0cm
\centerline{\rotatebox{0}{\scalebox{0.8}{\includegraphics{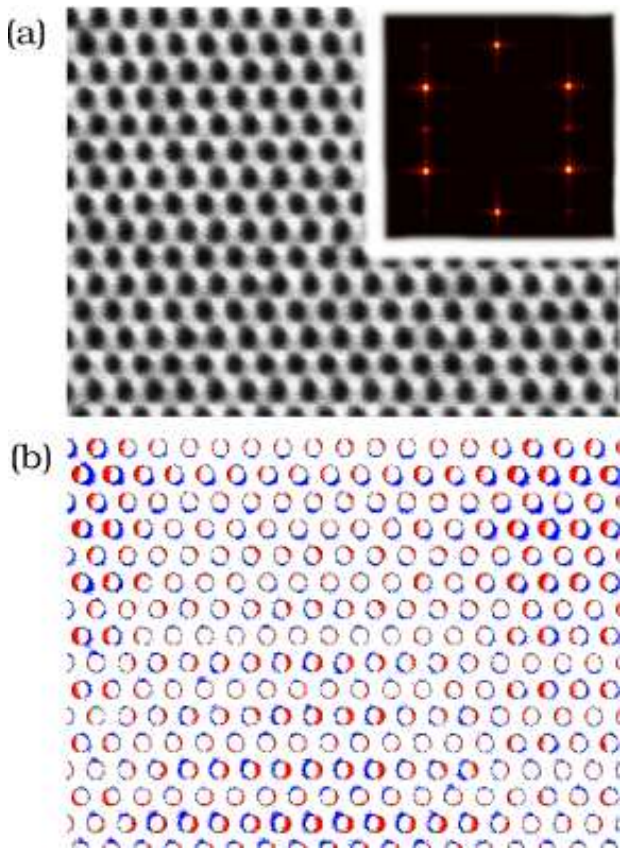}}}}
\caption{ (color online)
(a) Top-view SEM of a ZnO PhCS. The inset shows structural Fourier transform from the digitized SEM. Long-range periodicity is reflected in the six maxima at the positions corresponding to the perfect lattice. (b) Difference between the digitized SEM of real sample and the perfect honeycomb lattice reveals the structural disorder.}
\end{figure}

To investigate how the disorder affects the rates of vertical and lateral leakages of light from a PhCS, we consider a system schematically depicted in the inset of Fig. 2(a). A dielectric slab of thickness 180 nm and refractive index $n=2.35$ is sandwiched between air and substrate ($n_{sub}=1.78$). Within the slab, $N$ infinitely-long grooves run parallel to $y$-axis. The width of a groove is 22 nm, the lattice constant of the disorderless structure is 100 nm. We consider light propagating in $x$-$z$ plane, with the electric field along $y$-axis. Such system is 2D, that allows numerical simulation of large statistical ensembles of random systems. Despite the simplification, the system in Fig. 2(a) retains the property essential for our study of PhCS laser: the possibility of vertical (along $z$-axis) {\it and} lateral (along $x$-axis) radiative escape. Using the finite-difference time-domain (FDTD) method, we find the mode of the passive system that has the highest $Q_{tot}$ \cite{disphc}. A Gaussian pulse was launched at all spatial points in the slab and the energy is allowed to leak out radiatively. Simulation area is terminated by uniaxially perfectly matched absorbing layer that absorbs all outgoing waves. The pulse excites all modes within $30$ nm wavelength range around $400$ nm. After the initial multi-mode decay the field distribution is stabilized and the longest-lived mode can be seen. This is further confirmed by observing a mono-exponential decay of the total energy \cite{disphc,mode_delocalization} stored in the system that allows determination of $Q_{tot}$. By integrating Poynting vector over the corresponding interfaces\cite{mode_delocalization}, we obtained the outgoing flux in the vertical and horizontal directions, and $Q_{\perp}$ and $Q_{||}$. In our simulation, $Q_{tot}^{-1}=Q_{\perp}^{-1}+Q_{||}^{-1}$ relation was satisfied numerically to within $2\%$.

Fourier transform of the spatial profile of electric field at the interface between the slab and substrate gives the mode's distribution in $k_{||}$ (in-plane component of the wavevector) space. In a perfectly periodic structure, the bandedge mode has the highest-$Q_{tot}$. It is spatially extended in $x$, thus has a narrow distribution in $k_{||}$ [thick dashed curve in Fig. 2(a)]. Next we intentionally create a defect by increasing the spacing between two neighboring grooves at the center of the pattern to 150 nm. The highest-$Q_{tot}$ mode is localized around this artificial defect with a localization length of 140 nm. Strong localization in $x$ results in a broad distribution in $k_{||}$ [thin dashed curve in Fig. 2(a)], with the maximum lying closer to the edge of substrate light-cone [dash-dotted vertical line in Fig. 2(a)]. Its $Q_{tot}$ is limited by $Q_{\perp}$, which is about three times smaller than the corresponding $Q_{||}$ in a system of $N=24$. In contrast, the bandedge mode is concentrated well beyond the light-cone in $k_{||}$-space, thus its $Q_{\perp}$ is much higher. However, its spatial extension makes the lateral leakage larger, hence its $Q_{tot}$ is limited by $Q_{||}$.

\begin{figure}
\centerline{\rotatebox{0}{\scalebox{0.8}{\includegraphics{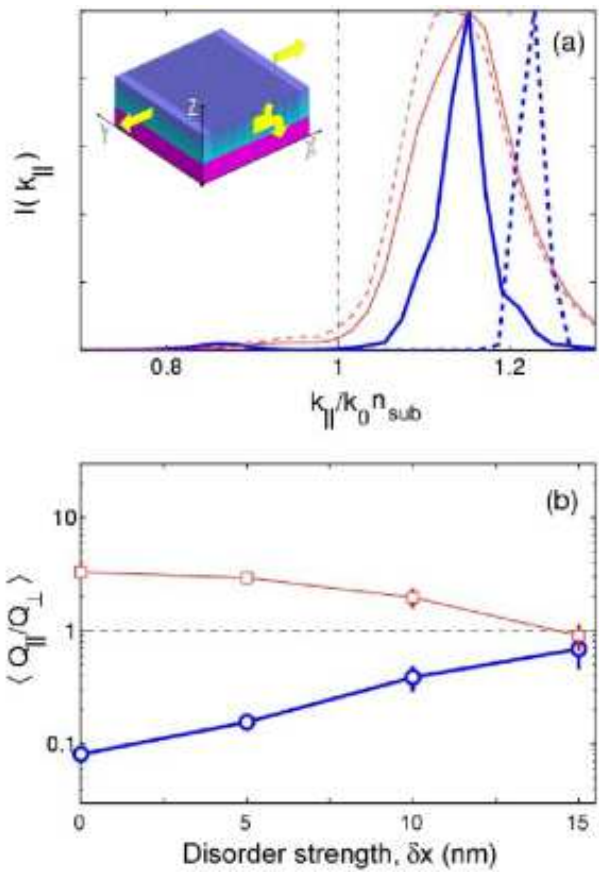}}}}
\caption{(color online)
(a) $k_{||}$-distributions of the highest-$Q_{tot}$ modes at one pixel beneath the slab/substrate interface. Thin/thick dashed curve represents the mode found in the disorderless system ($N=24$)  with/without an artificial defect. The corresponding solid curves are representative examples of the highest-$Q_{tot}$ modes in these systems with position disorder ($\delta x=10$ nm). The vertical line marks the substrate light-cone boundary. The inset is a schematic sketch of the simulated structure. Arrows show the directions of in-plane and out-of-plane radiative leakages. (b) Squares/circles represent $\langle Q_{||}/Q_{\perp}\rangle$, averaged over 300 random realizations of $N=24$ system with/without the artificial defect, versus disorder strength $\delta x$.}
\end{figure} 

To simulate the position disorder of air cylinders in real structure [Fig. 1(b)], random variation of groove position $x_n$ is introduced. We choose $\Delta x_n$ randomly from a uniform distribution with the standard deviation $\delta x=$5, 10, 15 nm. $\delta x$ characterizes the ``strength'' of disorder. As the disorder is introduced, the highest-$Q_{tot}$ state differs from realization to realization, and the correspondent $Q_{||}$, $Q_{\perp}$ as well as the frequency vary. We study statistical distributions of these parameters and their dependences on disorder strength $\delta x$ and system size $N$.

In small systems ($N$ = 12 and 24) with an artificial defect and weak disorder ($\delta x=5$ nm), the highest-$Q_{tot}$ modes always concentrate around the defect at the center of the pattern. These modes become more spatially extended than those without disorder. Therefore, their $k_{||}$-distribution is narrowed and $k_{||}$ component within the light-cone is significantly reduced [Fig. 2(a)]. This reduction leads to a decrease in the vertical leakage, thus, an increase in $Q_{\perp}$. Meanwhile, $Q_{||}$ starts increasing as the mode gets less localized in real space.  The ensemble-averaged $\langle Q_{||}/Q_{\perp}\rangle$, shown in Fig. 2(b), decreases monotonously to unity with increase of disorder strength. Therefore, {\it disorder removes the imbalance between vertical and lateral leakages} of a single defect state, making $\langle Q_{||}\rangle \sim \langle Q_{\perp} \rangle$. As a result, the ensemble-averaged quality factor  $\langle Q_{tot}\rangle$ is slightly higher than that without disorder.  In a larger system or with stronger disorder, the highest-$Q_{tot}$ mode is no longer pinned at the artificial defect. Instead, it can explore the entire pattern to find the optimum configuration for the best vertical and lateral confinement. This leads to a further increase of $\langle Q_{tot}\rangle$.

With the introduction of disorder, bandedge mode becomes less extended. As its ``tail'' moves away from the boundaries of the pattern, the lateral leakage decreases, thus $Q_{||}$ increases. Meanwhile, the distribution in $k_{||}$-space is broadened and shifted closer to the light-cone edge [Fig. 2(a)]. The increase in vertical leakage results in a decrease of $Q_{\perp}$. The ensemble-averaged $\langle Q_{||}/Q_{\perp}\rangle$, shown in Fig. 2(b), rises continuously to unity with increasing disorder strength. Again, disorder balances the  vertical and lateral leakages of the bandedge mode, as it does to the defect state. However, for a bandedge mode the increase in $\langle Q_{||}\rangle$ is not as large as the decrease in $\langle Q_{\perp} \rangle$, thus $\langle Q_{tot} \rangle$ is slighter lower than that without disorder. Nevertheless, as the pattern size $N$ increases, the total leakage rate decreases monotonically: $\langle Q_{tot}^{-1}\rangle \propto N^{-\alpha}$ [Fig. 3(a)]. The exponent $\alpha$ decreases from 2.3 at $\delta x=5$ nm to 1.9 at $\delta x=15$ nm. Even with the largest disorder we simulated ($\delta x=15$ nm), no saturation of $\langle Q_{tot}^{-1}\rangle$ with $N$ is observed up to $N=48$. This behavior differs fundamentally from that of a photonic crystal waveguide, where optical loss increases exponentially with its length. In contrast, a disordered PhCS laser {\it benefits} from an increase of the pattern size, simply because a larger system provides a bigger pool of modes from which the highest-$Q_{tot}$ mode can be selected. This effect should be more pronounced in PhCS microlasers with 2D periodicity [Fig. 1(a)], due to the larger phase-space compared to the numerically simulated systems with 1D periodicity.

\begin{figure}
\vskip 0cm
\hskip 0cm
\centerline{\rotatebox{-90}{\scalebox{0.3}{\includegraphics{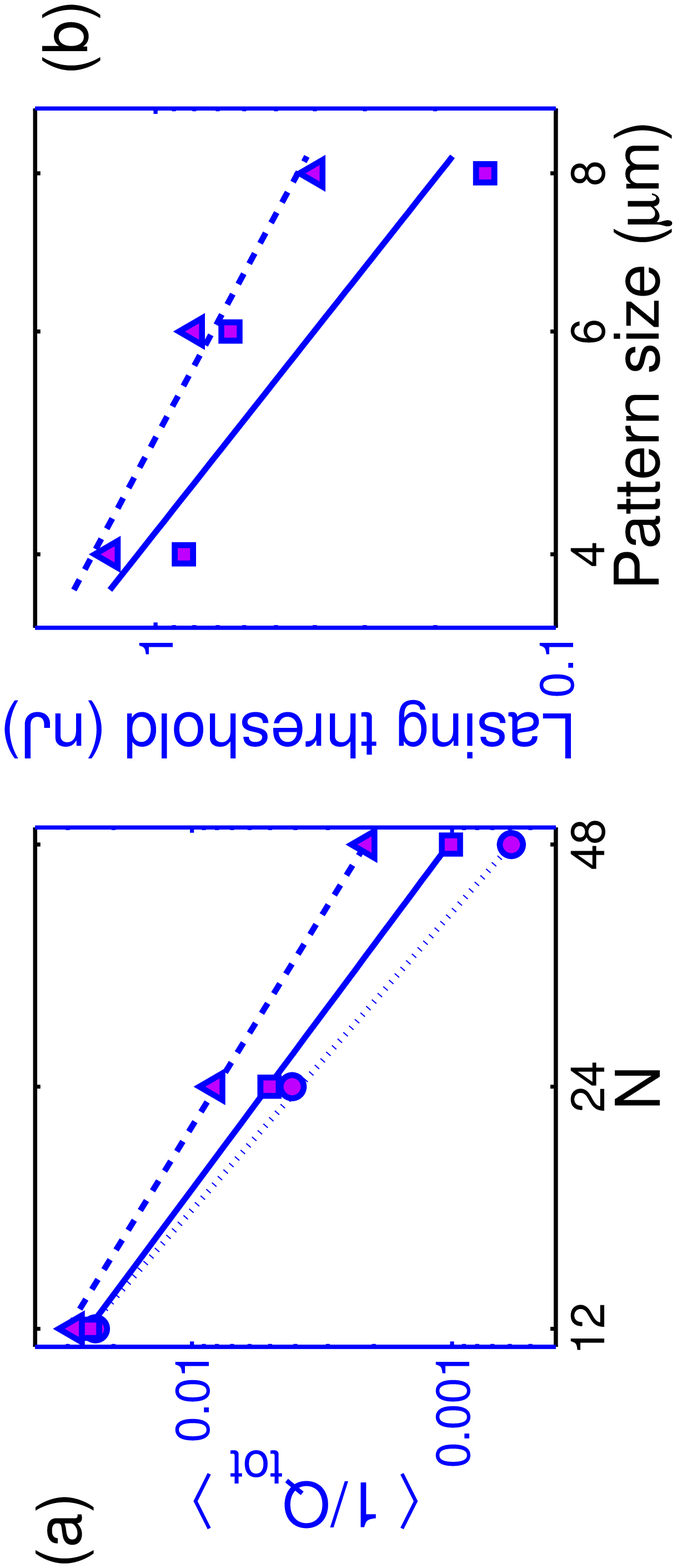}}}}
\vskip 0.5cm
\centerline{\rotatebox{0}{\scalebox{0.8}{\includegraphics{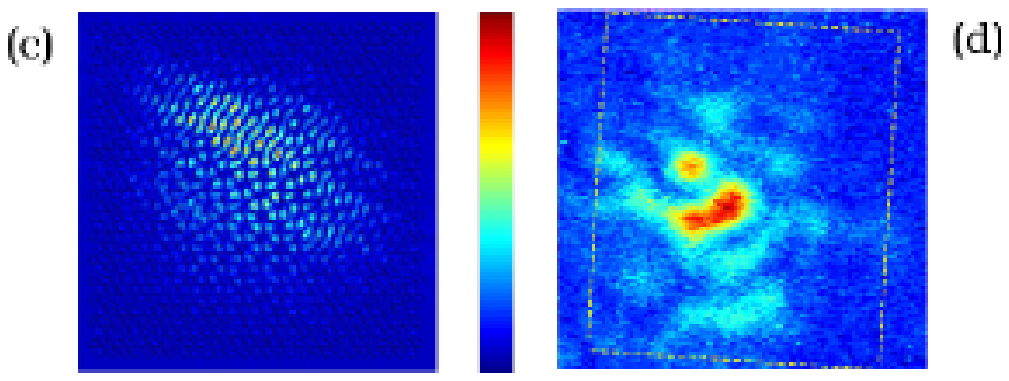}}}}
\caption{(color online)
(a) $\langle 1/Q_{tot}\rangle$ for the highest-$Q_{tot}$ modes found numerically in PhCS depicted in the inset of Fig. 2(a). The average is taken over an ensemble of $300$ random realizations. The squares and triangles represent the results for $\delta x=5$ nm and $15$ nm, respectively. Circles correspond to the disorderless system. Dotted, solid and dashed lines are $N^{-\alpha}$ fits with $\alpha=2.7,\ 2.3$ and $1.9$ respectively. (b) Ensemble-averaged incident pump pulse energy at the lasing threshold, measured in the samples of $a=115$ nm (squares) and $a=118$ nm (triangles), versus the pattern size. The data are fitted with the power law dependence as in (a) with $\alpha=2.5$ (solid line) and $1.7$ (dashed line). (c) Intensity profile of the highest-quality mode found numerically in a 2D disordered hexagonal array of air holes in the dielectric with refractive index $n=1.55$. The pattern size is $5\times 5$ $\mu$m, lattice constant $a=186$ nm, and the hole radius $R=65$ nm. The wavelength of the mode is $393$ nm. It is located at the dielectric bandedge, and the electric field is concentrated inside the dielectric. (d) Measured intensity distribution of the lasing mode in a ZnO PhCS with $a=115$ nm and $R =0.25a$. Dashed box outlines the boundary of the $8\times8$ $\mu$m pattern.}
\end{figure}

Experimentally, we fabricated ZnO PhCS of dimensions $4\times 4$ $\mu$m, $6\times 6$ $\mu$m, and $8\times 8$ $\mu$m [Fig. 1(a)]. Since the complete photonic bandgap in ZnO PhCS without ``undercut'' was quite narrow \cite{our_pbg_calc}, it was technically challenging to overlap PBG with ZnO gain spectrum. By adjusting the magnification of focused ion beam system, we were able to change the lattice constant $a$ in fine steps of 3 nm over a wide range $100-160$ nm. The ratio of the air hole radius $R$ to the lattice constant $a$ was also varied from $0.20$ to $0.30$. In this way, we could tune PBG continuously through ZnO gain spectrum. We also introduced an artificial defect by missing an air hole. Structural analysis as in Fig. 1(b) gives the average displacement of a hole $\delta r \simeq 0.22 R$.

ZnO PhCS was optically pumped by the third harmonics of a pulsed Nd:YAG laser ($\lambda$ = 355 nm, 10 Hz repetition rate, 20 ps pulse width) at room temperature \cite{our_apl}. In $8\times 8$ $\mu$m patterns without intentionally-introduced structural defect, the ensemble-averaged lasing threshold exhibited a pronounced minimum at $a=115-130$ nm and $R=0.25 a$ [Fig. 4(a)]. To understand this phenomenon, we calculated the photonic bands in ZnO PhCS using the computational technique described in Ref.\cite{our_pbg_calc}. The frequency dependence of ZnO refractive index was taken into account. In Fig. 4 (b), the wavelength of the dielectric bandedge $\lambda_d$ for the fundamental PBG of TE modes \cite{our_apl} is plotted against the lattice constant $a$. The structural parameters were extracted from the SEM of our samples. The ZnO slab thickness $t = 180$ nm, and $R/a= 0.245$. By comparing the lasing wavelength to $\lambda_d$ in Fig. 4(b), we confirmed that the lasing modes were located in the vicinity of the dielectric bandedge. This can be explained by two factors: (i) the electric field of the modes near the dielectric bandedge is concentrated inside ZnO, and thus experience more gain [Fig. 3(c)]; (ii) the vanishing group velocity at the bandedge enhances light amplification \cite{gain_at_be}. The dip in the measured lasing threshold [Fig. 4(a)] is attributed to spectral overlap of the dielectric bandedge with ZnO gain spectrum. In Fig. 3(b), the measured lasing threshold decreases monotonously with the pattern size for $a$ = 115 nm and 118 nm.  These data agree qualitatively with the numerical simulation results shown in Fig. 3(a). In all patterns with intentionally missed air holes, the lasing modes were not pinned at the location of the missing hole due to the existence of better confined modes away from the defect. This observation is in line with our numerical simulation of large patterns with single artificial defect. 

\begin{figure}
\vskip 0cm
\centerline{\rotatebox{0}{\scalebox{0.37}{\includegraphics{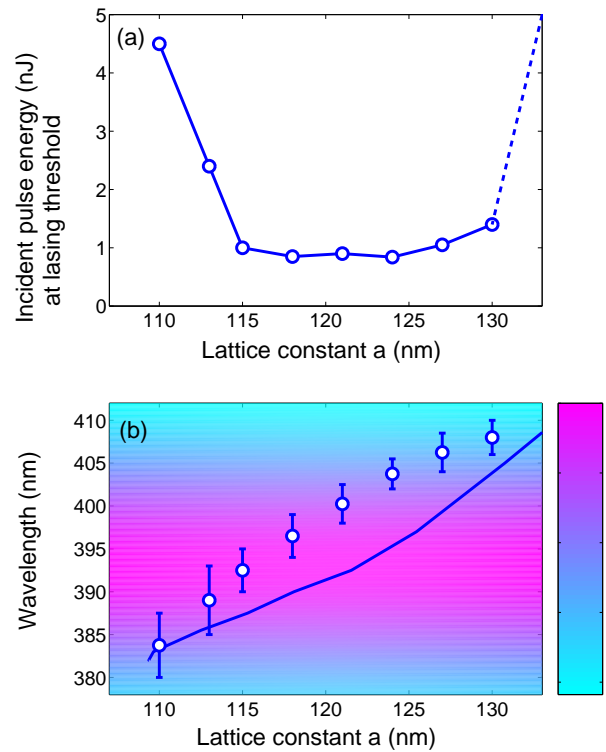}}}}
\caption{(color online)
(a) Experimentally measured incident pump pulse energy at the lasing threshold (averaged over five to ten samples) as a function of lattice constant $a$. (b) The circles are the measured mean wavelength of lasing modes, the error bar depicts the range of lasing wavelengths. The solid curve represents the wavelength of the calculated dielectric bandedge $\lambda_d$ in ZnO PhCS as a function of $a$. The shade of the background qualitatively describes the position and width of ZnO gain spectrum.}
\end{figure}

In summary, the structural disorder may lead to self-optimization of optical confinement in a PhCS and formation of high-$Q_{tot}$ modes which serve as the lasing modes. In a sufficiently large  PhCS with short-range disorder, a microcavity with balanced $Q_{\perp}$ and $Q_{||}$ can be formed spontaneously without any carefully-designed structural defects. Despite the disorder, photonic bandedge effect enables us to efficiently extract optical gain and to fine-tune the lasing wavelength from $383$ nm to $407$ nm with sample-to-sample fluctuation of about $5$ nm in ZnO PhCS lasers. 

This work was supported by the National Science Foundation under the grant no. ECS-0244457.

\end{document}